\documentclass[pra,twocolumn,superscriptaddress,longbibliography]{revtex4-1}
\usepackage{amssymb,amsmath,amsfonts,bm}
\usepackage{graphicx}
\usepackage{epstopdf}
\usepackage{times}
\usepackage{float}
\usepackage{lipsum}
\usepackage{color}

\usepackage{hyperref}
\hypersetup{  colorlinks=true, linkcolor=blue, citecolor=red, urlcolor=blue  }

\begin{document}

\title{Coherent Control with User-Defined Passage}

\author{Bao-Jie Liu}
\affiliation{School of Physics, Harbin Institute of Technology, Harbin 150001, China}
\affiliation{Department of Physics, Southern University of Science and Technology, Shenzhen 518055, China}
\author{Man-Hong Yung}  \email{yung@sustech.edu.cn}
\affiliation{Department of Physics, Southern University of Science and Technology, Shenzhen 518055, China}
\affiliation{Shenzhen Institute for Quantum Science and Engineering, Southern University of Science and Technology, Shenzhen 518055, China}
\affiliation{Guangdong Provincial Key Laboratory of Quantum Science and Engineering, Southern University of Science and Technology, Shenzhen 518055, China}
\affiliation{Shenzhen Key Laboratory of Quantum Science and Engineering, Southern University of Science and Technology, Shenzhen,518055, China}

\date{\today}

\begin{abstract}
 Stimulated Raman adiabatic passage (STIRAP) is a standard technique to combat experimental imperfections and can be used to realize robust quantum state control, which has many applications in physics, chemistry, and beyond. However, STIRAP is susceptible to decoherence since it requires long evolution time. To overcome this problem, stimulated Raman “user-defined” passage~(STIRUP) is proposed, which allows users to design the passages unlike the STIRAP but fast and robust against both decoherence and experimental imperfections. Here, we further develop a more general STIRUP method. Comparing with shortcut to adiabaticity and its' variants, the generalized STIRUP is more simpler and compatible with more complex energy-level structure and many-body systems. Furthermore, the generalized STIRUP has many important applications such as geometric phase measurement, coherent population transfer, and quantum state preparation. Specifically, as examples, we show how to realize the high-fidelity quantum state transfer and entangled state generation in a robust way via STIRUP with the state-of-the-art experimental superconducting circuits.

\end{abstract}


\maketitle

\section{Introduction}
Coherent control of quantum states is of fundamental importance in quantum information processing, such as high-precision sensing~\cite{Kasevich2002,Kotru2015}, robust quantum computation~\cite{Farhi2001,Monroe2001}, and simulation~\cite{Kim2010}. There are a lot of quantum optimal control methods developed to realize high-fidelity state preparations that are robust against decoherence and control parameter variations~\cite{DD1,DD2,DD3,Kaveh,BB1,SUPCODE,Vasilev2009,Bergmann1998,Vitanov2017}.
One of the more effective methods is the adiabatic control protocol, where control parameters are slowly changed to avoid transitions between different sets of eigenstates. In particular, stimulated Raman adiabatic passage (STIRAP)~\cite{Bergmann1998} based on the ``adiabatic passage" without a lossy excited state has become a standard coherent control technique. STIRAP has two attractive properties~\cite{Vitanov2017} which are (i) robust against loss due to spontaneous emission from the loss excited state; (ii) robust against small variations of experimental control parameters, such as pulse amplitude,duration and phase. 
With these two properties, STIRAP has been demonstrated for realizing quantum state control~\cite{Bergmann2019} and constructing holonomic quantum gates~\cite{Zanardi1999,Duan2001}, respectively. 
However, these applications are limited by the intrinsic adiabatic condition, which implies long evolution time. Therefore the schemes based on STIRAP would no doubt be susceptible to the environment-induced decoherence~\cite{Vitanov2017}, especially, dephasing error. More specifically, the performance of STIRAP-based schemes will be greatly reduced in the solid-state quantum systems such as superconducting circuit~\cite{Premaratne2017,Kumar2016,Xu2016,Vepsalainenarxiv2017}, quantum dot~\cite{Koh2013} and NV center in diamond~\cite{Bason2012NatP} since the dephasing error is the main source of decoherence in these platforms.
 
To overcome such a problem, recent work~\cite{Niu2019}, called STIRUP, demonstrates that the notion of the adiabatic passage in STIRAP can be extended without the adiabatic condition, which determines the driving pulses directly from the inverse engineering of some ``user-defined" passage in a three-level system. With the flexibility of STIRUP, one can optimize different objectives for different tasks, such as minimizing leakage error, enhancing robustness against control errors, speeding up quantum control. However, there exists no efficient way of constructing these for complex systems, such as many-level and many-body system.

Here, we present a general scheme for STIRUP, which allows users to design fast and robust passages against the decay, dephasing and imperfection for many-level and many-body system.  
Moreover, our general scheme can be regarded as a generalization of STIRAP and stimulated Raman shortcut-to-adiabatic passage (STIRSAP)~\cite{Du2016NatC,Torrontegui2013,D2019,Unanyan1997,Emmanouilidou2000,Demirplak2003,Demirplak2005,Demirplak2008,Berry,chenxi,Song2016,Lewis1969,chenxi2011,Laforgue2019,Gue2019,Baksic,Zhou2016np,Huang2017,Yan2019,Liu2019PRL,Dorier2017}, as it yields identical results to STIRAP in the adiabatic condition, as shown in Fig.~(\ref{setup}a).  Our approach is more robust against dephasing noises than STIRAP, and is more simpler and efficient to speed up STIRAP than traditional STIRSAP without relying on counteradiabatic driving~\cite{Unanyan1997,Emmanouilidou2000,Demirplak2003,Demirplak2005,Demirplak2008,Berry,chenxi}, dynamical invariant~\cite{Lewis1969,chenxi2011,Laforgue2019,Gue2019} and dressed-state~\cite{Baksic,Zhou2016np,Huang2017}.  It should be note that the scheme~\cite{Laforgue2019} use the dynamical invariant method with optimal control to find an exact passage with `` $\Lambda$" system for robust quantum state transfer. However, the dynamical invariant requires the dynamical symmetry of the system, which is difficult to find in some complicated system swith  $N\ge4$ ( here \emph{N} denote the  energy level)~\cite{D2019}. We also note that the work~\cite{Dorier2017} using parameterized states for solving nonlinear Schr\"{o}dinger equation for nonlinear quantum systems (like BEC), the authors did not consider the optimizations of the population of the intermediate state and robustness against control error for liner systems. 

For the applications of the general scheme, we consider an $N$-level pod ($N$-pod) system for realizing robust many-qubit state transfer and entangled state generation via STIRUP. with the state-of-the-art superconducting experimental parameters, we performed numerical simulation to show that the entangled Bell state and W state can be achieved with the high fidelities 99.7\% and 98.4\%, respectively.

\section{General model of STIRUP}

The family of quantum control problems investigated in this work is focused on the problems of preparing a certain $N$-dimensional target state $|\psi_{T}\rangle=\sum_{n=1}^{N} c_{n}|n\rangle$, starting from a given initial state $|\psi_{0}\rangle=\sum_{n=1}^{N} b_{n}|n\rangle$. One is required to find the corresponding driving Hamiltonian $H(t)$ for achieving such a goal. For this purpose, the term ``passage" may be defined as a parameterized time-dependent state $|\phi_{P}(t)\rangle$ satisfying boundary conditions at time $t=0$ and $t=T$: $|\phi_{P}(t=0)\rangle = |\psi_{0}\rangle$, and $|\phi_{P}(t=T)\rangle|=|\psi_{T}\rangle $.

An example is the application of STIRAP on a three-level $\{|1\rangle ,|2\rangle ,|3\rangle\}$ lambda system~\cite{Bergmann1998,Vitanov2017,Bergmann2019}. The ``adiabatic" passage is defined by $\left|\phi_{A}(t)\right\rangle \equiv \cos \theta(t)|1\rangle-\sin \theta(t)|2\rangle$, where $\theta(t)$ is required to satisfy the boundary conditions: $\theta(0)=0$, and  $\theta(T)=\pi/2$. In this way, one can achieve the goal of population transfer from $|1\rangle$ to $|2\rangle$. The passage is adiabatic because the strategy of STIRAP is based on the design of a driving Hamiltonian $H(t)=h_{13}(t)|1\rangle\langle 3|+h_{23}(t)|2\rangle\langle 3|+h.c.$, where the passage becomes an eigenstate of $H(t)$ when $\tan\theta(t)=h_{13}(t)/h_{23}(t)$. Then, it is sufficient to vary the driving amplitudes $\Omega_{0,1}$ sufficiently slowly compared with the energy gap.  

\begin{figure}[tbp]
\centering\includegraphics[width=8.5cm]{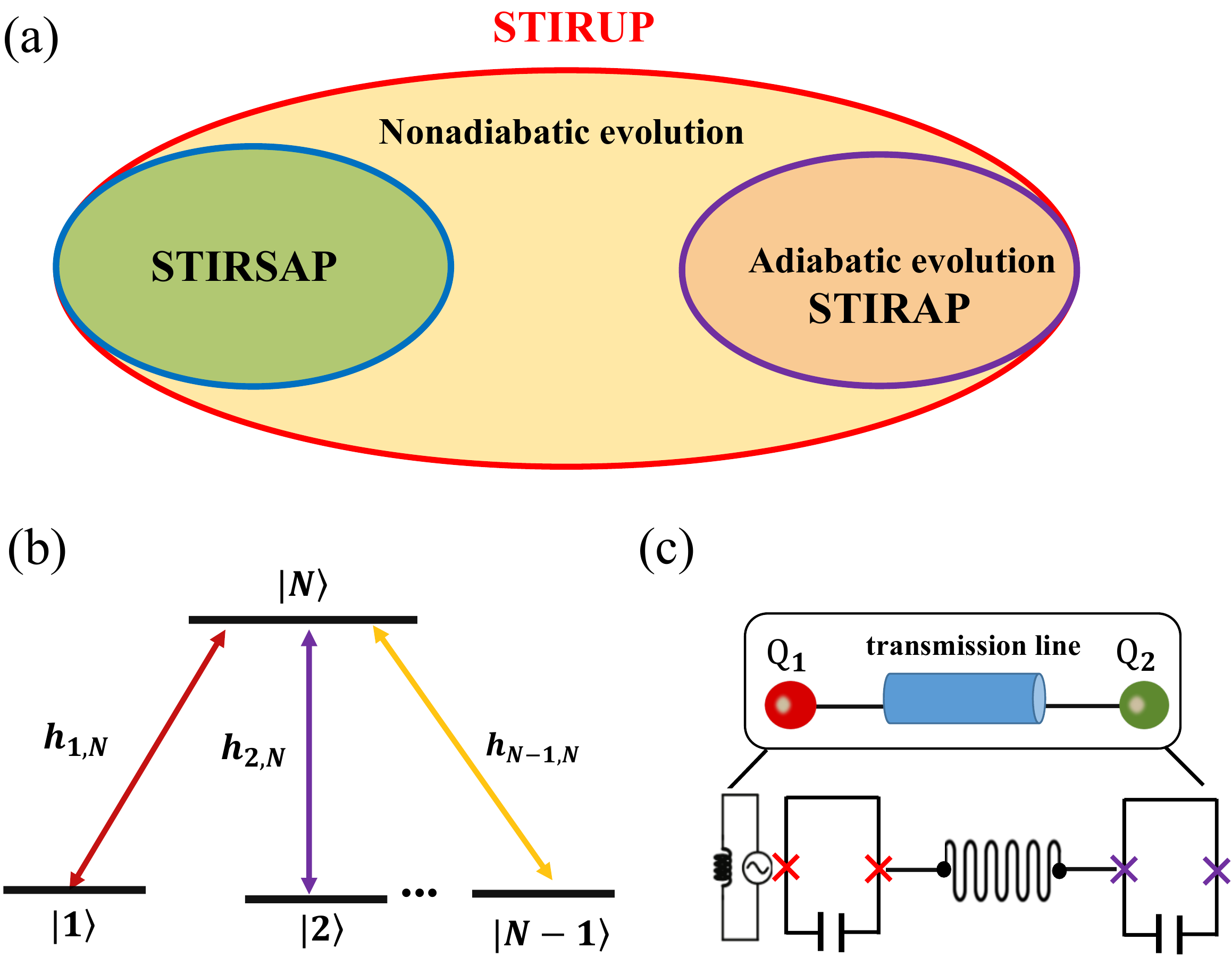}\caption{\label{setup} The illustration of our proposed implementation. (a) Schematic of relations of known quantum control methods, STIRSAP and STIRAP are included in STIRUP. 
(b) The $N$-pod level structure and coupling configuration. (c) Schematic of our circuit consisting of two capacitively coupled qubits, where $Q_{2}$ is biased by an ac magnetic flux to periodically modulate its transition frequency. $Q_{1}$ and $Q_{2}$  are coupled via transmission-line-resonator.}
\end{figure}

However, the notion of the adiabatic passage in STIRAP can be extended without the adiabatic condition. Specifically, we choose to determine the driving pulses directly from the inverse engineering of some ``user-defined" passage $|{{\phi_P}\left( t \right)}\rangle$. Without loss of generality, the “user-defined" passage $|{{\phi_P}\left( t \right)}\rangle$ can be generally parameterized as
\begin{equation}\label{STATE}
| {{\phi_{P}}\left( t \right)}\rangle=\sum_{n=1}^{N} a_{n}(t)|n\rangle \ ,
\end{equation}
where the time-dependent coefficients ${a_{n}(t)}$ satisfy boundary conditions $a_n(0)=b_n$, $a_n(T)=c_n$ and normalized condition $\sum_{n=1}^{N} |a_{n}(t)|^{2}=1$ . 

From the Schr\"{o}dinger equation, the time dependence of the driving pulses are governed by the following set of equations, $\sum_{n} h_{mn}(t)a_{n}(t)= i\dot{a}_{m}(t)$.
However, unlike the traditional approach of solving the Schr\"{o}dinger equation, where the Hamiltonian is usually given, here both the matrix elements $h_{mn}(t)=\langle m|H(t)| n\rangle$, and the coefficients $a_{n}(t)$ have to be determined consistently. In general, one may reduce the degrees of freedom of the passage by requiring it to evolve along a certain pathway, and the Hamiltonian would naturally has some physical constraints. Therefore, solving the set of coupled equations may not necessarily be a trivial task. The point is that the state preparation problem constraints only the user-defined passage through the different boundary conditions, but the trajectory can be designed for optimizing additional objectives, such as noise robustness, decorherence errors, or time duration, as discussed below. Recently a three-level state transfer via STIRUP has been experimentally implemented in a superconducting circuits~\cite{Niu2019}, with the transfer fidelity significantly improved by STIRUP.

As a demonstration, we consider an $N$-level pod ($N$-pod) system with an $N$-dimensional state space, where a single level labeled $|N\rangle$ is coupled to $N-1$ levels labeled $|m\rangle (m=1,...,N-1)$ as shown in Fig.~(\ref{setup}b). Previously, STIRAP was applied to 3-level systems~\cite{Bergmann1998} for realizing an efficient coherent population transfer via adiabatic passage $|\phi_{A}(t)\rangle$ from an initial state $|1\rangle$ to a target state $|2\rangle$, which is achieved by means of a two-photon process involving the driving pulses fields $h_{13}(t)$ and $h_{23}(t)$. 
Afterwards, STIRAP was extended to an $N$-pod system for realizing adiabatic population transfer~\cite{Ivanov2006,Amniat2011} and simulating non-Abelian gauge fields~\cite{Dalibard2011,Barnett2012}. Here, with essentially the same physical setting, we apply STIRUP pulses for quantum control of $N$-pod system, avoiding the adiabatic constraint.

The corresponding $N$-pod Hamiltonian $H(t)$, is described by~\cite{Vitanov2017}
\begin{equation}
H(t)=\sum_{m=0}^{N-1} \frac{\omega_{m N}}{2} \sigma_{z_{m}}+h_{mN}(t) \cos \left(\omega_{d_{m}} t\right) \sigma_{x_{m}}
\end{equation}
where $\omega_{m N}$ is the energy splitting of the level $|m\rangle$ and $|N\rangle$. and $\sigma_{z_{m}}=|m\rangle\langle m|-| N\rangle\langle N|$ and $\sigma_{x_{m}}=|m\rangle\langle N|+| N\rangle\langle m|$ 
is the Pauli matrix defined by the ground state $|m\rangle$ level and excited level $|N\rangle$; and the control field $h_{mN}(t)$ drives the $|m\rangle\leftrightarrow|N\rangle$ transitions. When the resonant condition $\omega_{mN}=\omega_{d_{m}}$ is satisfied, under the rotating-wave approximation and the interaction picture, the system Hamiltonian can be written as,
\begin{equation} \label{H1}
H_{I}(t)=\sum_{m=1}^{N-1}h_{mN}(t)\sigma_{x_{m}},
\end{equation}
Here, our “user-defined” passage for a $N$-pod system can be parameterized as:
\begin{align}\label{Parameter}
a_{1}& = \cos \gamma \prod_{i=1}^{N-2} \cos \chi_{i},\quad a_{N-1} = -\cos \gamma \sin \chi_{N-2} \\ \nonumber \
a_{k}& = \cos \gamma \sin \chi_{k-1} \prod_{i=k}^{N-2} \cos \chi_{i},\quad a_{N} = -i \sin \gamma  \ .
\end{align}
where $\chi_{i}(t)$ and $\gamma(t)$ are generally time-dependent variables to be determined below with the integer $k$ ranging from $k\in (1,N-1)$. 
To achieve state preparation  from $|\phi_{P}(0)\rangle = |1\rangle$ to $|\phi_{P}(T)\rangle = |\psi_{T}\rangle$, it is sufficient to impose the following boundary conditions: $\gamma(0)=\gamma(T)=0$, $\chi_{i}(0)=0$ and $\chi_{i}(T)=s_{i}$.

\begin{figure*}[tbp]
\centering
\includegraphics[width=14cm]{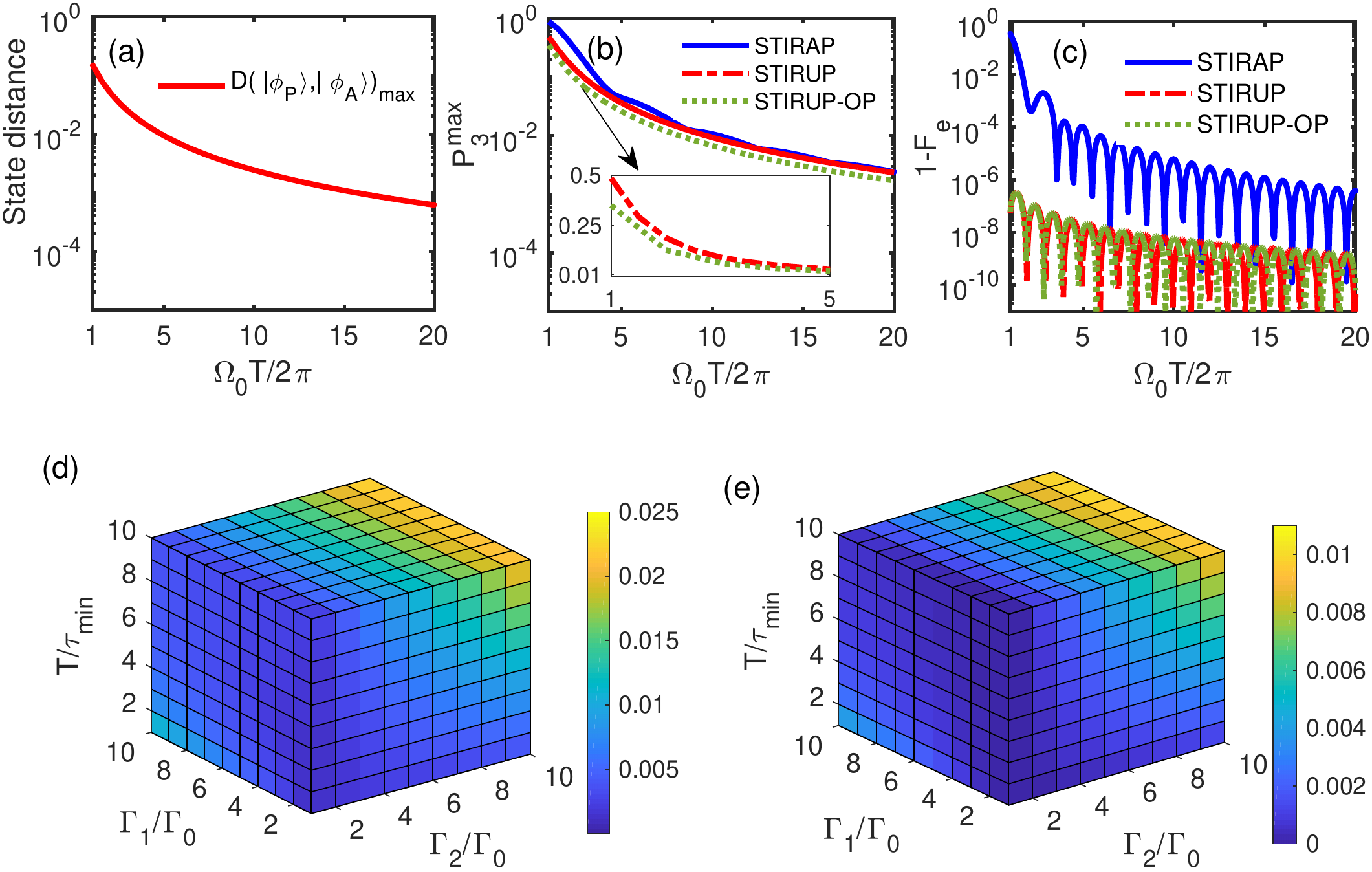}
\caption{(a) The maximum state distance  between the “user defined" passage $|\phi_{P}\rangle$ and  the adiabatic passage $|\phi_{A}\rangle$ as a function of dimensionless parameter $\Omega_{0}T$ (b) The maximum population $P^{max}_{3}$ in the intermediate level $|3\rangle$ of the STIRUP, STIRUP-OP and STIRAP. (c) The error transfer efficiency as the functions of $\Omega T$ without consideration the decoherence effect. The error transfer efficiency of (d) STIRUP and (e) STIRUP-OP as the functions of the decay rate $\Gamma_{1}$, the dephasing rate $\Gamma_{2}$ and the evolution time $T$. }\label{TWO}
\end{figure*}

Substituting Eq. (\ref{Parameter}) and Eq. (\ref{H1}) into the Schr\"{o}dinger equation, the inverse engineering control fields are given by 
\begin{align}\label{NTRIOD}
h_{1 N} = &f_{1}\prod_{i=1}^{N-2} \cos \chi_{i} \\ \nonumber \
h_{k N} =&\left[f_{l}\sin \chi_{m-1}-\dot{\chi}_{m-1} \cos \chi_{m-1} \cot \gamma\right] \prod_{i=m}^{N-2} \cos \chi_{i} \\ \nonumber \
h_{N-1 N}&=\dot{\chi}_{N-2} \cot \gamma \cos \chi_{N-2}-\dot{\gamma} \sin \chi_{N-2}
\end{align}
where $f_{x}\equiv\left(\dot{\gamma}+\cot \gamma \sum_{l=x}^{N-2} \dot{\chi}_{l} \tan \chi_{l}\right)$. Note that extra care must be taken at the initial time $t = 0$, as the boundary conditions would imply a divergence of the driving pulses whenever $\left.\cot \gamma\right|_{\gamma \rightarrow 0} \rightarrow \infty$. To overcome such a problem, we enforce additional boundary conditions:  $\dot{\chi}_{m}(0)=\dot{\chi}_{m}(T)=0$, to maintain the combination $h_{mN}(t)$ to be finite.

Under the coupled differential equations in Eq. (\ref{NTRIOD}), we can realize arbitrary state preparation using our STIRUP by choosing proper boundary conditions.  For example, to realize coherent population transfer from initial quantum state $|\psi_{0}\rangle=|1\rangle$ to the target state $|\psi_{t}\rangle=|N-1\rangle$, the boundary conditions are set to $\chi_{i}(T)=s_{i}=\pi/2$. One possible set of solution is found to be  
\begin{eqnarray}\label{BETA}
\begin{aligned} \gamma(t)=&\arctan{\left[\frac{\dot{\chi}(t)}{\Omega_{0}}\right]} \\ 
\chi_{i}(t)=&\chi(t)=\arctan{\left[\frac{1-\cos(\pi t/T)}{1+\cos(\pi t/T)}\right]}
\end{aligned}
\end{eqnarray}
where $\Omega_{0}$ is a constant to control the pulse amplitude. Similar to previous dressed-state method~\cite{Baksic}, we can also replace the parameter $\Omega_{0} \rightarrow \Omega_{0}\left[1+Q\left(1-\cos \frac{2 \pi t}{T}\right)^{4}\right]$ to reduce $\gamma$ and suppress the population of intermediate level $|N\rangle$ ($P_{N}(t)=|\sin\gamma(t)|^{2}$). The time-independent parameter $Q$ can be numerically optimized to minimize the population in intermediate state for each operation time without increasing the maximum Rabi strength. One simple method to search $Q$ is the brute force search optimisation with a random initial $Q$. Note that we can avoid or further suppress the excitation of intermediate state but at the cost of increasing the maximum Rabi strength (see Fig. \ref{TWO}(b) for details).
 
In this way, for the case $N=3$, the time dependence of Rabi control pulses $h_{13}(t)$ and $h_{23}(t)$ can also be determined numerically. More specifically, using the Eq. (\ref{Parameter}), the “user-defined" passage is taken as $\left|\phi_{P}(t)\right\rangle=\cos\chi(t)\cos\gamma(t)|1\rangle-\sin\chi(t)\cos\gamma(t)|2\rangle-i\sin\gamma(t)|3\rangle$, which is corresponding to $a_{1}(t)=\cos\chi(t)\cos\gamma(t)$, $a_{2}(t)=\sin\chi(t)\cos\gamma(t)$ and $a_{3}(t)=-i\sin\gamma(t)$.
Consequently, the time dependence of the control pulses
can be determined by Eq. (\ref{NTRIOD}) as $ h_{13}(t) =\dot{\chi} \cot \gamma \sin \chi(t)+\dot{\gamma}(t) \cos \chi(t)$ and $h_{23}(t)=\dot{\beta}(t) \cot \gamma(t) \cos \chi-\dot{\gamma}(t) \sin \chi(t)$.  Therefore, the nonadiabatic state transfer from $|1\rangle$ to $|2\rangle$ can be realized.

In general, we have many methods to choose the variables $\gamma(t)$ and $\chi(t)$ satisfying the equation Eq. (\ref{NTRIOD}) and the boundary conditions, which makes it possible for our method to be compatible with most of the optimization schemes, such as GRAPE~\cite{Khan2005}. Specifically, we can choose the proper parameters of $\gamma(t)$ and $\chi(t)$ with the modified sine Fourier series~\cite{Leo2017}, i.e., $\gamma(t) =2 \chi(t)+\sum_{n=1} C_{n} \sin [2 n \chi(t)]$ and $\chi(t)=\frac{\pi t}{2T}+\sum_{m=1} S_{m} \sin \left(2 m \frac{\pi t}{T}\right)$ for high-fidelity and robust state transfer under different noises, where the series of resulting coefficients $S_{m}$ and $C_{n}$ can be numerically determined via GRAPE~\cite{Khan2005}.

\section{The generality and superiority of STIRUP}

Here, we demonstrate that our STIRUP is a general protocol that can reproduce the outcomes of all other methods by designing different passages, and take the STIRAP as an example. 
Before that, a minimum time $\tau_{min}$ of STIRUP is defined by the constraint that the STIRUP pulse cannot exceed its maximal amplitude of STIRAP at each moment, $\left\{h_{13}(t), h_{23}(t)\right\}_{STIR UP}^{\max } \leq\left\{h_{13}(t), h_{23}(t)\right\}_{STIRAP}^{\max }$, due to the limitation of experimental conditions.  For the solution in Eq.~(\ref{BETA}), we numerically get $\tau_{min}=3.24/\Omega_{0}$. When the above adiabatic condition is satisfied $|\dot{\theta}| \ll 1$, the dark state $\left|E_{0}(t)\right\rangle$ is an approximate solution of Schr\"{o}dinger equation, i.e., $\left|E_{0}(t)\right\rangle=| {{\Phi_{p}}\left( t \right)}\rangle$. Therefore the STIRAP scheme can be viewed as one STIRUP passage under the adiabatic condition.  To further illustrate it, we gradually increase the operation time $T$ from $T=\tau_{min}$ to $T=40\tau_{min}$ to be close to “local adiabatic condition", where adiabatic condition is usually well satisfied under the condition, $\Omega_{0}T>2\pi\times10$, obtained from experience and numerical simulation studies~\cite{Bergmann1998}. 

Here, we use  the maximum state distance defined by $D\left(\left|\phi_{P}\right\rangle,\left|\phi_{A}\right\rangle\right)_{\max }=\frac{\left|T_{r}\left(\left|\phi_{P}\right\rangle\left\langle\phi_{P}|-| \phi_{A}\right\rangle\left\langle\phi_{A}\right|\right)_{\max }\right|}{2}$, to evaluate the gap between STIRAP and STIRUP. From the Fig.~~\ref{TWO}(a), the result can clearly verify our analysis that STIRUP gradually becomes STIRAP as the local adiabatic condition approaches.  Comparison of the maximum population $P^{max}_{3}$ in the intermediate level $|3\rangle$ of the STIRUP to STIRAP, STIRUP has better performance than STIRAP regardless of whether the adiabatic condition is met as shown in Fig.~\ref{TWO}(b). In addition, we plot the error transfer efficiency of STIRAP, STIRUP and STIRUP-OP as the function of the ``adiabatic parameters" $\Omega_{0}T$ in Fig.~\ref{TWO}(c). Furthermore, to further reduce the intermediate state occupation against decoherence, we can numerically design the optimal parameter $Q$ by brute force search optimisation. The optimization results are shown in Table \ref{table1}, where we have bounded the initial guess of $Q$ as $Q\in[-0.1,0.1]$. In Fig. \ref{TWO}(d), we plot the error transfer efficiency $1-F_{e}$ as a function of $\Omega_{0}T$ for STIRAP, STIRUP and STIRUP-OP, where $F_{e}$ is transfer efficiency defined by
$F_{e}=\left|\left\langle\phi_{P}(T) | \psi_{t}\right\rangle\right|^{2}$.

The decoherence process  is unavoidable, and understanding its effects is crucial for quantum state control. The performance of STIRUP, STIRUP-OP and STIRAP can be simulated by using the Lindblad master equation~\cite{Lindblad}
\begin{eqnarray}  \label{Master}
\dot\rho(t) &=& i[\rho(t), H(t)]  +  \frac{\Gamma_{1}}{2} \mathcal{L}(S_{-}) + \frac{\Gamma_{2}}{2} \mathcal{L}(S_{+}),
\end{eqnarray}
where $\rho$ is the density matrix of the considered system, $\mathcal{L}(A)=2A\rho_1 A^\dagger-A^\dagger A \rho_1 -\rho_1 A^\dagger A$ is the Lindbladian of the operator $A$, $S_{-}=|1\rangle\langle 3|+| 2\rangle\langle 3|$, $S_{+}=|2\rangle\langle 2|+| 3\rangle\langle 3|$; $\Gamma_{1}$ and  $\Gamma_{2}$ are the decay and dephasing rates of the system, respectively. In our simulation, we plot the error transfer efficiency of STIRUP, STIRUP-OP and STIRAP with different the decoherence rates $\Gamma_{1}$ and $\Gamma_{2}$ with the unit of $\Gamma=2\pi\times5$ kHz, and  the different evolution time T as shown in Fig.~\ref{TWO}(d) and ~\ref{TWO}(e). We can clearly see that our STIRUP and STIRUP-OP model can give the optimal evolution time for different $\Gamma_{1}$ and $\Gamma_{2}$ of different experimental parameters.

\begin{table}[htb]
\centering \caption{Optimal pulses with the control parameters $Q$ for different evolution time $T$.}
\begin{tabular}{cccccccccc}
\hline
$T/\tau_{min}$  & 1 & 2 & 3 & 4 & 5 & 6 & 7 & 8 & 9\\
\hline
Q/100  & 2 & 1.58 & 1.37 & 1.26 & 1.2 & 1.16 & 1.14 & 1.12 & 1.11\\
\hline
$T/\tau_{min}$  & 11 & 12& 13 & 14 & 15 & 16 & 17 & 18-39 & 40 \\
\hline
Q/100  & 1.09 & 1.09 & 1.09 & 1.08& 1.08 & 1.08 & 1.08 & 1.07 & 1.06 \\
 \hline
\end{tabular}\label{table1}
\end{table}

 \begin{figure}[htb]
\centering
\includegraphics[width=8.5cm,height=7.0cm]{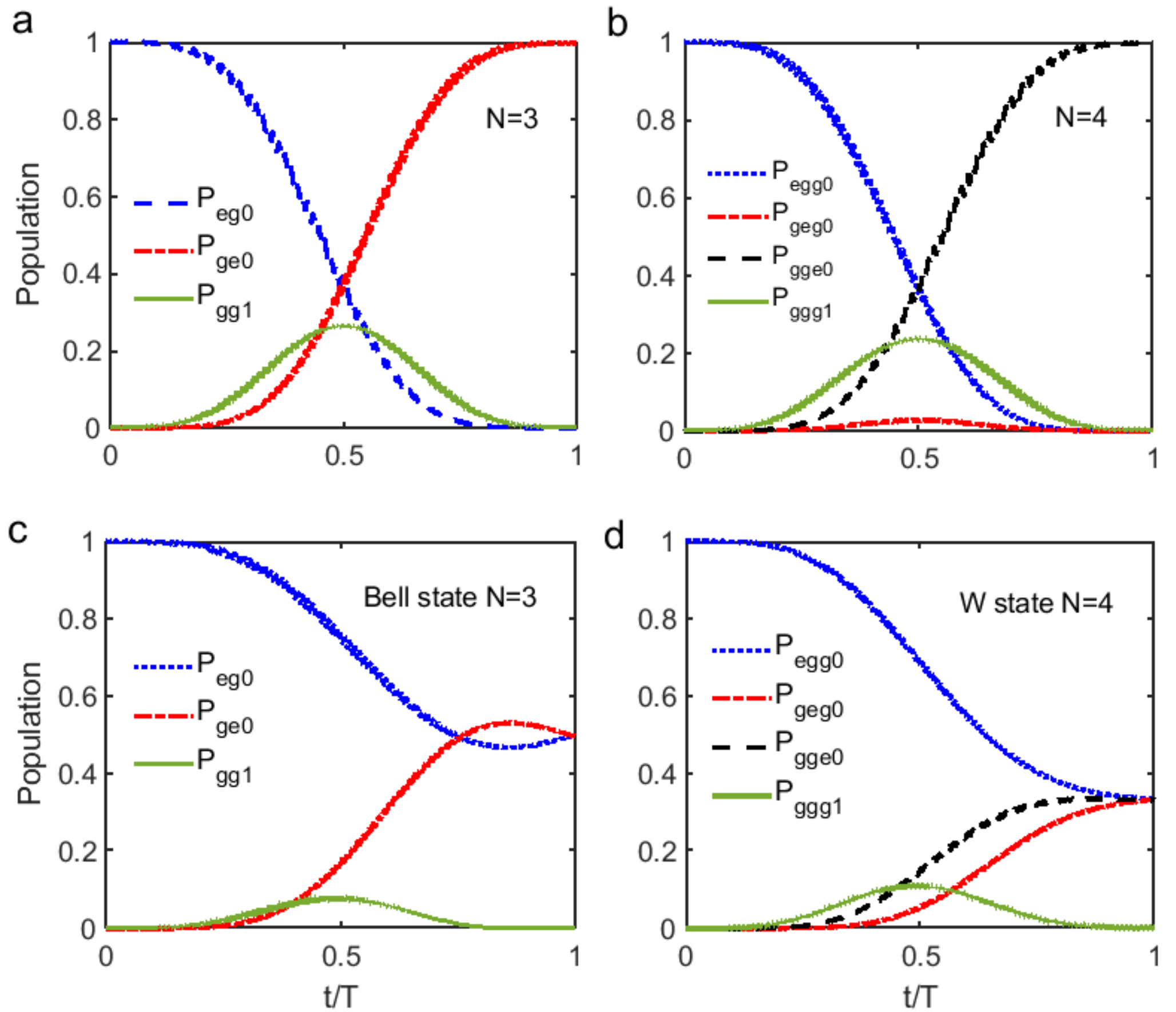}
\caption{ The dynamical population of (a) two-qubit and (b) three-qubit QST, and the dynamical population of (c) Bell state and (d) W state via STIRUP.}\label{population}
\end{figure}

\section{Application of STIRUP on superconducting circuits}
In this part, we will make our discussion explicit by demonstrating its application in realistic systems, namely, superconducting circuits. Specifically, we shall focus on a many-qubit superconducting quantum processor in Ref.~\cite{Song2017,Xu2018,SongC2019}, as shown in Fig.~(\ref{setup}d). All qubits are interconnected by a central cavity (bus resonator), and the frequency of each qubit can be individually manipulated with its control lines.
In the rotating wave approximation and ignoring the cross talk between qubits, the Hamiltonian of the system~\cite{Tavis1968} is given by
\begin{equation}\label{SCS}
H_{3}(t)=\omega_{c} a^{+} a+\sum_{j=0}^{N-1}\left[\frac{\omega_{q_{j}}}{2}\sigma^{z}_{j}+g_{j}\left(\sigma_{j}^{+} a+\sigma_{j}^{-} a^{+}\right)\right]\ ,
\end{equation}
where $\omega_{c}$ is the frequency of cavity, $\omega_{q_{j}}$ is the energy splitting of the qubit, $\sigma_{j}^{z}$ is the Pauli matrix of the $j$th qubit in its eigenbasis,  $g$ is the qubit-cavity coupling strength, $\sigma_{j}$ ($\sigma^{+}_{j}$) is the qubit lower (raising) operator, and $a^{+}$ ($a$) is the creation (annihilation) operator of cavity. To obtain tunable coupling between the two qubits, we add an ac magnetic flux on the $j$th qubit to periodically modulate its frequency as
$\omega_{q_{2}}(t)=\omega_{q_{2}}+\varepsilon_{j}(t) \cos \left(\nu_{j} t\right)$,
where $\varepsilon_{j}(t)$, and $\nu_{j}$ are the modulation amplitude, and frequency, respectively. Moving into the interaction picture, the effective interaction Hamiltonian is
\begin{equation}\label{EHA}
H_{4}(t)=\sum_{j=0}^{N-1}\widetilde{g}_{j}(t)\left(\sigma_{j}^{+} a+\sigma_{j}^{-}a^{+}\right).
\end{equation}
where the time-dependent effective coupling is $\widetilde{g}_{j}(t)=g_{j}\emph{J}_{1}(\varepsilon_{j}(t))$, and $\emph{J}_{1}$ is the Bessel function.

Here, we firstly demonstrate the state transfer from the initial state $|eg0\rangle$ to the target state $|ge0\rangle$ in superconducting quantum processor. In a single-excitation subspace spanned by $\left\{|eg0\rangle,|ge0\rangle,|gg1\rangle\right\}$, denoting the states of qubits and the cavity, the Hamiltonian Eq. (\ref{EHA}) has the same form as the Eq. (\ref{H1}) of N-pod system corresponding to N=3.

As shown in Fig. (\ref{setup}c), for the tripod system with N=4 ($j$=4), the passage is chosen as 
\begin{equation}
\begin{aligned}\left|\phi_{t}(t)\right\rangle =&{\cos \beta_{1} \cos \gamma_{1}\cos\phi_{1}|egg0\rangle}+\sin\beta_{1}\cos \gamma_{1}|gge0\rangle \\ &+ \cos\beta_{1}\cos\gamma_{1}\sin\phi_{1}|geg0\rangle-i \sin \gamma_{1}|ggg1\rangle\end{aligned}\ .
\end{equation}
The boundary condition of $\beta_{1}$ and $\gamma_{1}$ is the same as $\beta_{0}$ and $\gamma_{0}$ for the case N=3. In addition, the control parameter $\phi_{1}$ satisfies the condition $\phi_{1}(0)=0 (\pi)$, the corresponding solution is given by $\beta_{1}(t)=\beta_{0}$, $\gamma_{1}(t)=\gamma_{0}$, and $\phi_{1}(t)=\arctan{\left(\frac{\cos\beta_{0}\cos\gamma_{0}-1}{\cos\beta_{0}\cos\gamma_{0}+1}\right)}$.
According to the Eq. (\ref{NTRIOD}), we get the Rabi pulse shapes of $\widetilde{g}_{0}(t)$,  $\widetilde{g}_{1}(t)$ and $\widetilde{g}_{2}(t)$. Therefore, the high-fidelity QST can be realized from the initial state $|egg0\rangle$ to the target state $|gge0\rangle$.

Numerical simulation of the QST population dynamics are shown in Fig. (\ref{population}a) and (\ref{population}b), where the high-fidelity QST fidelities of for the three-qubit and four-qubit QST can be obtained with 99.4\% and 99.19\% using the following set of the current experimental parameters~\cite{Barends2013,Barends2014,Chen2016}: the cavity and qubit frequency is $\nu_{j}=\omega_{c}-\omega_{q_{j}}=2\pi\times1$ GHz, the gate time is set T = 82 ns, and the dissipation parameters of the cavity and qubits are taken $\Gamma_{1,2}=\Gamma_{C}=\Gamma$.

Secondly, we can realize a high-fidelity entangled state preparation via one-step STIRUP.  To generate two-qubit Bell state $|\Psi_{B}\rangle=\frac{|eg0\rangle+|ge0\rangle}{\sqrt{2}}$, the boundary conditions are setted as  $\gamma_{0}(0)=\gamma_{0}(T)=0 (\pi)$ and $\beta_{0}(0)=0$, $\beta_{0}(T)=\pi/4$,with the corresponding parameters $\beta(t)=\frac{\pi t}{2T}$ and $T=90$ ns. As shown in Fig. (\ref{population}c), with the numerical simulation, the Bell stat is formed with fidelity as high as 99.7\% due to loss excited state  $|gg1\rangle$ population and short-time evolution. To further improve the fidelity, we can combine STIRUP with the quantum optimal control pulse to minimize the leakage caused by the higher levels~\cite{Xue2016PRA,Rol2019}. Similar to the approach of generation Bell state, the tripartite entanglement W state $|\Psi_{w}\rangle=\frac{|egg0\rangle+|geg0\rangle+|gge0\rangle}{\sqrt{3}}$ can be generated by choosing $j=4$, $\beta(t)=S_{1}t/T$, $\gamma_{1}=1/\sqrt{3}\sin S_{1}$ and $T=98.5$ ns with $S_{1}=0.5678\pi$. As shown in Fig. (\ref{population}d),  the W state $|\Psi_{w}\rangle$ can be achieved with high fidelity 98.4\%. The genuine entanglement states of Bell state and W state violate entanglement witnesses that rule out bi-separability,  have been generated. This ability to couple three-qubit system and create entangled states which are qualitatively different is a significant step towards salable quantum information processing with superconducting devices. 

\begin{figure}[tbp]
\centering
\includegraphics[width=8.5cm,height=7.0cm]{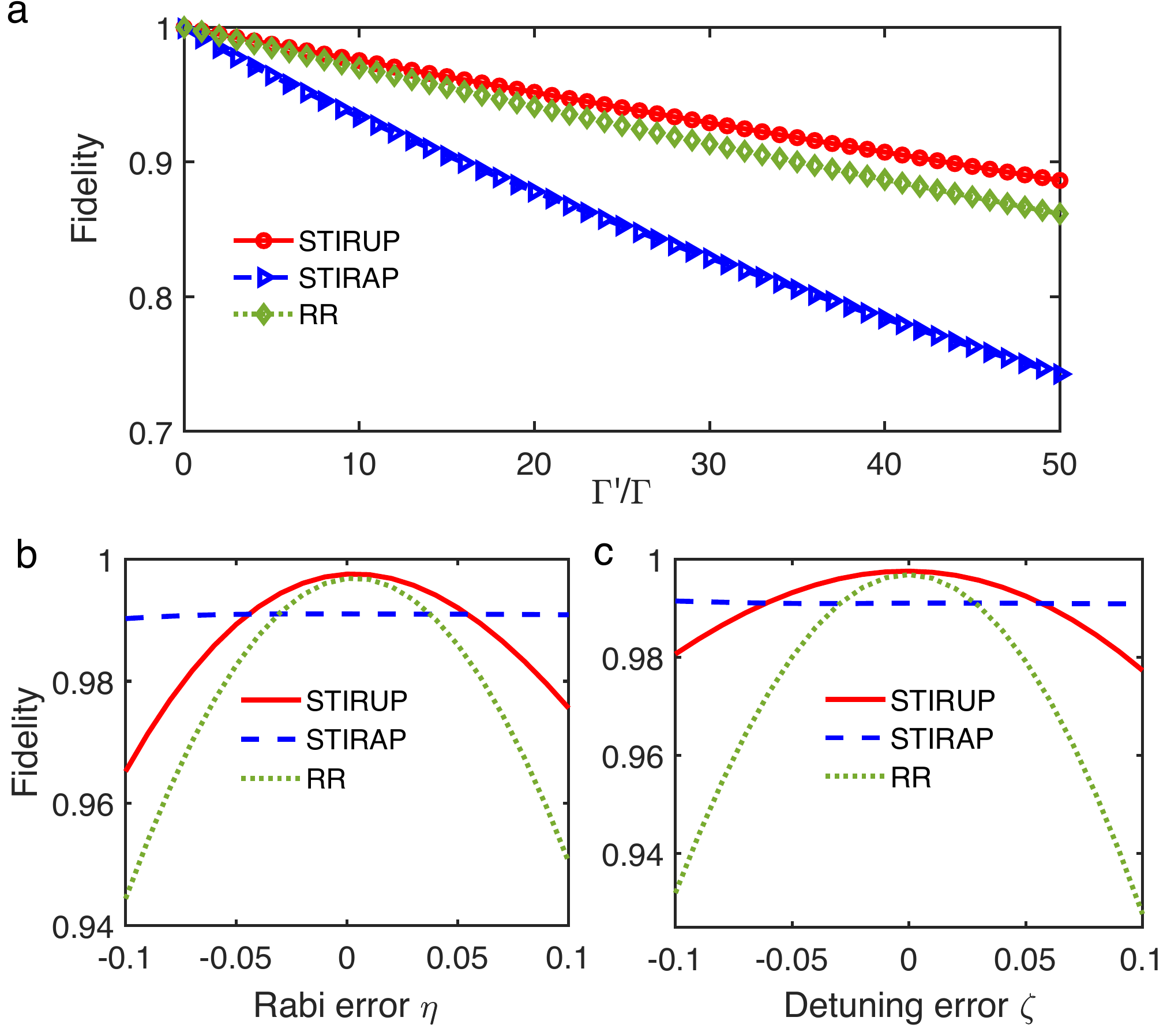}
\caption{(a) The two-qubit QST fidelities of STIRUP, STIRAP and Rabi Resonate (RR) as a function of (a) decoherence rates $\Gamma^{'}$ (in unit of $\Gamma$), (b) Rabi error $\eta$ and (c) detuning error $\zeta$.}\label{single}
\end{figure}

\section{Robustness}
We proceed to show the superiority of STIRUP on robustness against environmental noises and experimental imperfections.  Firstly, to investigate the robustness against decoherence comparing with STIRAP~\cite{Chang2020} and resonate Rabi pulses (RR) in Ref.~\cite{Mlynek2012,Chapman2016,Egger2019,Li2018} in a three-state system. We plot the QST fidelity defined by $\mathrm{F}=\left|\left\langle\phi_{I} | \phi_{t}(T)\right\rangle\right|^{2}$ as functions of the decoherence rate $\Gamma^{'}$ (in unit of $\Gamma$), where $|\phi_{I}\rangle$ represents the ideal target state for STIRUP, STIRAP, and RR schemes, as shown in Fig. (\ref{single}a). Clearly, STIRUP is more robust against the decoherence effect comparing with STIRAP and RR schemes due to its' short evolution time and loss excited-state population.

Secondly, to test the robustness of STIRUP against experimental pulse errors caused by an usual slow quasistatic noise, we add a static deviation to the strength of driving pulse, i.e., ${\widetilde{g}_{max}}\to (1 + \eta ){\widetilde{g}_{max}}$, where $\eta\in[-0.1,0.1]$ represents the Rabi error. In other words, the Hamiltonian becomes $H_{3}(t) \to (1 + \eta )H_{3}(t)$. Comparing our STIRUP with the STIRAP and RR methods, we simulated the performance of QST with the same pulse error under the decoherence effect. As shown in Fig. (\ref{single}b), STIRUP is not only always more robust than the STIRAP scheme but also RR scheme when the Rabi error is small.  
Finally, we further investigate the sensitivity of the STIRUP protocol to the variation of the qubit frequency detuning $\omega_{q_{j}} \to \omega_{q_{j}}+\delta$, where we denotes the detuning error as $\delta=\zeta\widetilde{g}_{max}$ with $\zeta\in[-0.1,0.1]$. As shown in Fig. (\ref{single}c), the transfer fidelity of STIRUP is also more robust against the detuning error $\zeta$ than others within the small detuning error. Moreover, our STIRUP does has advantages over STIRAP in terms of robustness with the Rabi error $\eta\approx\pm 0.02$ and detuning error $\zeta\approx0.03$ in the recent experiment of STIRAP~\cite{Chang2020}.

\section{Conclusion}
We have presented a general STIRUP scheme for complex systems that allows users to design different fast and robust passages against both decoherence and imperfection by directly engineering solutions of the Schr\"{o}dinger equation. Consequently, this approach is simpler and more efficient when extending to many-level and many-qubit system comparing with shortcuts to adiabaticity. Furthermore, the general STIRUP has many important physical applications such as geometric quantum computation, coherent population transfer, and quantum state preparation. Specifically, we realized many-level state transfer and many-qubit entangled state generation with high fidelity and noise robustness by using STIRUP.


\acknowledgments
We also thank Prof. S.-L. Su and Dr. X.-M. Zhang for valuable discussions. This work is supported by the Natural Science Foundation of Guangdong Province (Grant No. 2017B030308003), the Key R \& D Program of Guangdong province (Grant No. 2018B030326001), the Science, Technology and Innovation Commission of Shenzhen Municipality (Grant No. JCYJ20170412152620376 and No. JCYJ20170817105046702 and No. KYTDPT20181011104202253), National Natural Science Foundation of China (Grant No. 11875160 and No. U1801661), the Economy, Trade and Information Commission of Shenzhen Municipality (Grant No. 201901161512), Guangdong Provincial Key Laboratory (Grant No.2019B121203002).


\end{document}